\documentclass[aps,prd,nofootinbib]{revtex4}
\usepackage{graphicx}
\usepackage{epsfig}

\begin{document}

\title{Radial acceleration relation and dissipative dark 
matter}

\author{Olga~Chashchina}
\email{chashchina.olga@gmail.com}
\affiliation{\'{E}cole Polytechnique, Palaiseau, France}

\author{Robert~Foot}
\email{rfoot@unimelb.edu.au}
\affiliation{ARC Centre of Excellence for Particle Physics at the
Terascale,
School of Physics, University of Melbourne, Victoria 3010 Australia}

\author{Zurab~Silagadze}
\email{Z.K.Silagadze@inp.nsk.su}
\affiliation{ Budker Institute of 
Nuclear Physics and Novosibirsk State University, Novosibirsk 630
090, Russia }

\begin{abstract}
Observations indicate that ordinary matter, the baryons, influence the 
structural properties of dark matter on galactic scales. One such
indication 
is the radial acceleration relation, which is a tight correlation
between the 
measured gravitational acceleration and that expected from the baryons.
We show here that the dark matter density profile that has been
motivated by 
dissipative dark matter models, including mirror dark matter, can
reproduce 
this radial acceleration relation.
\end{abstract}

\maketitle

\section*{Introduction}

The  matter content of the Universe can be inferred from the cosmic
microwave 
background and large scale structure studies to be dominated by a
nonbaryonic 
dark matter component \cite{cmb}. Dark matter has also been inferred to
exist 
on galaxy scales, with rotation curve measurements providing the most
detailed 
information about the structural properties of dark matter on small
scales, 
see e.g. \cite{rot0a,rot0b,nat,rot1,rot2,rot3,rot4}.
From such studies evidence that the dark matter structural properties 
are influenced by ordinary matter, baryons, is in abundance. The first
such 
indication was the Tully-Fisher relation \cite{tf}, which relates galaxy 
luminosity with the asymptotic value of the rotational velocity.
Another such indication is the radial acceleration relation, 
which is a tight correlation between the measured gravitational
acceleration 
and  that expected from the baryons \cite{8} (for earlier closely
related work
see \cite{3}). See also e.g. \cite{btf,s1,s2,s3} for other related
studies and 
discussions of the baryonic influence of the structural properties of
dark 
matter in galaxies.

The radial acceleration relation is a one-parameter
formula which summarizes many of the empirical correlations that
rotation
curve measurements show with the baryon content. Such a relation was in
fact
initially motivated by modified Newtonian 
dynamics (MOND) \cite{mond},  which was originally posed as an
alternative to 
dark matter. 
While the radial acceleration relation arises quite naturally within
MOND 
it might seem unclear how such a connection with baryons could arise
in dark matter theories.
There is however a kind of dark matter which requires nontrivial
interactions 
with baryons, called dissipative dark matter
\cite{fv,10,11,12,13,14,15}.  
Further, we shall show here that the dark matter density profile, that
has been 
motivated by dissipative dark matter models, can reproduce 
the radial acceleration relation.

%The empirical correlations that rotation curve measurements show
%with baryon content can be explained with a relatively simple
%one-parameter 
%formula within the MOND framework. Dark matter theories {\it must}
%reproduce 
%the radial acceleration relation and other galactic scaling relations
%(within measured uncertainties), 
%if they are to describe nature.

Given that the radial acceleration relation  was originally introduced
in 
MOND, let us briefly digress to indicate how it arises in that
framework.
MOND postulates that the acceleration $g$ of a test mass in a galaxy 
gravitational field is fully determined by the Newtonian  acceleration 
expected at the same location due to the gravitational field strength 
$g_{bar}$ created by the baryonic matter of the galaxy. In spherical or
axial 
symmetric case, the relation has the form 
\begin{equation}
g=g_{bar}\,\nu\left(\frac{g_{bar}}{a_0}\right),
\label{eq1}
\end{equation}
where $a_0$ is a new fundamental constant and the interpolating function 
$\nu(y)$ satisfies $\nu(y) = 1$ in the $y=g_{bar}/a_0\gg 1$ limit of
large 
accelerations (Newtonian limit). The behaviour of the interpolating
function
in the opposite limit $y\ll 1$ of small accelerations (deep-MOND regime) 
can be uniquely specified if the gravitational dynamics becomes scale 
invariant \cite{6,7}. Under the scale transformation $(t,\vec{r})\to
\Lambda\,
(t,\vec{r})$, the acceleration $g$ and the gravitational field strength 
$g_{bar}$ scale as follows:
\begin{equation}
g=\left |\frac{d^2\vec{r}}{dt^2}\right|\to\frac{1}{\Lambda}\,g,\;\;\;
g_{bar}\propto \frac{1}{r^2}\to\frac{1}{\Lambda^2}\,g_{bar}.
\label{eq2}
\end{equation}
Therefore, in order to have scale invariant dynamics in the limit 
$y \ll 1$, the interpolating function $\nu(y)$ must satisfy
\begin{equation}
\frac{1}{\Lambda}\,\nu\left(\frac{y}{\Lambda^2}\right)=\nu(y),\;\;\;y\ll
1.
\label{eq3}
\end{equation}
The solution of this functional equation is $\nu(y)=k/\sqrt{y}$ 
with some constant $k$. It is clear that this dimensionless constant $k$
can be absorbed in the definition of the critical acceleration $a_0$,
so that we can set $k=1$ without loss of generality.

The scale invariant gravitational dynamics at low accelerations is
perhaps 
the most important aspect of MOND, which (typically) does not  depend 
significantly on the specific form of the interpolating function
$\nu(y)$.
A simple choice for the interpolating function is \cite{7A}:
\begin{equation}
\nu(y)=\frac{1}{1-e^{-\sqrt{y}}} \ .
\label{eq4}
\end{equation}
%and
%^\begin{equation}
%a_0=(1.20\pm 0.02\pm 0.24)\times 10^{-10}~\mathrm{m}\,\mathrm{s}^{-2}.
%\label{eq5}
%\end{equation}
The relation Eq.(\ref{eq1})  with this functional choice for the
interpolating 
function was analyzed in the recent comprehensive study of \cite{8}
which 
involved 153 rotationally supported galaxies from the SPARC data base 
\cite{spc}. Within observational uncertainties this relation was found
to hold
in rotationally supported galaxies. The value for $a_0$ was estimated to
be: 
$a_0=(1.20\pm 0.02\pm 0.24)\times 10^{-10}~\mathrm{m}\,\mathrm{s}^{-2}$.
The results of \cite{8} have already generated considerable interest and
attempts 
to explain them, e.g.\cite{ref2,ref3,ref4}.

\section*{Dissipative dark matter model}

The observed correlations of rotation curve shapes with baryon content
in 
galaxies, for which the radial acceleration relation is an indicator, is
an 
interesting challenge for dark matter theories. This seems to be
particularly 
relevant to the often considered case of dark matter consisting of 
collisionless (or weakly interacting) particles. In particular, the 
correlations are observed to hold in gas-rich dwarf irregular galaxies
which 
are dark matter dominated at all radii. It is hard to envisage how
baryons, 
which are gravitationally insignificant, could have such a large
influence on 
the structural properties of the dark matter, if gravity were the only 
interaction with ordinary matter.  

Dissipative dark matter is a specific kind of dark matter candidate that 
actually requires nontrivial interactions with baryons for a consistent 
picture of dark matter in rotationally supported galaxies 
\cite{fv,10,11,12,13,14,15}. Such generic model involves 
considering a dark sector  consisting of dark matter particles coupled
with 
massless dark photons \cite{12}. A theoretically constrained case is
mirror 
dark matter \cite{16} which involves a duplicate set of particles and
forces 
with exactly the same fundamental parameters (particle masses and
coupling 
constants) as the standard particles and forces. In other words, we are 
envisaging dark matter with either similar or exactly the same particle 
properties to ordinary matter. 

Such dark matter is dissipative. We are interested in the parameter space
where 
dissipation plays an important role, that is, in the absence of any heat 
source the cooling rate is sufficiently strong for the galactic dark
matter to 
collapse into a disk on timescales much smaller than the Hubble time.
However, in the presence of heating a dissipative dark matter halo
could, in 
principle, exist in an approximately spherical pressure supported dark
plasma.
This is the scenario that we consider.\footnote{The only other scenario
involving
dissipative dark matter  that we are aware of
is the ``double disk'' dark matter model of \cite{15A,15B}. In that
model,
dissipative dark matter is
assumed to be only a subcomponent of the total nonbaryonic dark matter
sector. 
No heat source is envisaged in that model
so that the dissipative dark matter collapses forming a dark disk.}

In such models the only viable heat source identified is ordinary 
type-II
supernovae, which can become an energy source for the dark sector if
small 
kinetic mixing interaction exists. The kinetic mixing interaction 
\cite{fhe,holdom}
\begin{eqnarray}
{\cal L}_{int} = - \frac{\epsilon}{2}F^{\mu \nu} F_{\mu \nu}^{'} \ ,
\label{1x}
\end{eqnarray}
facilitates halo heating by enabling ordinary supernovae to be a source
of 
these `dark photons' \cite{raf,17}. These dark photons can transport a
large 
fraction of a supernova's core collapse energy to the halo (potentially
up to 
$\sim 10^{53}$ ergs per supernova for kinetic mixing of strength 
$\epsilon \sim 10^{-9}$). The supernovae generated dark photons 
propagate out into the halo where 
they can eventually be absorbed via some interaction process, with dark 
photoionization being the mechanism in the specific models studied  
\cite{10,11,12,13}.\footnote{
If halo heating is indeed due to ordinary supernovae then at early
times, 
prior to the onset of star formation, the dissipative dark matter might
have 
been able to cool to form a dark disk composed of dark stars made almost 
entirely of dark matter. In the particular case of mirror dark matter
such 
`mirror stars' can evolve extremely rapidly \cite{ber} and thereby
potentially 
source the heavy mirror nuclei (mirror metals) needed for the halo to
absorb 
the dark photon radiation from ordinary supernovae.
The remnant dark stars themselves are assumed to be a subdominant mass 
component of the halo at the present time.
}$\;$\footnote{Dwarf spheroidal and elliptical galaxies are devoid of
baryonic gas 
and have negligible active star formation. Ordinary supernovae cannot
provide 
a viable heat source, and thus a very different picture of the dark
matter 
halo in these galaxy types is envisaged. In these galaxies the dark
matter 
halo might have cooled and collapsed into compact objects, `dark stars'
in 
the case of mirror dark matter.}

Such strongly interacting dark matter can be modelled as a fluid
governed 
by Euler's equations. The halo is dynamic and is able to expand and
contract 
in response to the heating and cooling processes. It is envisaged that
for 
a sufficiently isolated galaxy its halo can evolve into a steady state 
configuration which is in hydrostatic equilibrium, and heating and
cooling 
rates are locally balanced. These conditions determine the temperature
and 
density profiles of the dark matter halo around such disk galaxies.
More generally, for interacting disk galaxies undergoing perturbations
or 
other non-equilibrium conditions (such as rapidly changing star
formation 
rate as in starburst galaxies) the dark matter density and temperature 
profiles would be time dependent and would require the full solution of 
Euler's equations.

Insight into the steady-state solution, applicable to an isolated disk
galaxy 
with a stable star formation rate,  arises from the following 
simple argument  \cite{12,13}. The (supernova sourced) heating rate at 
a particular location  
${\bf r}$ in the halo is proportional to the product of the dark matter 
density and dark photon energy flux at that point: $\Gamma_{heat}({\bf
r})
\propto n({\bf r})  F_{\gamma_D} (\bf{r})$.
The halo is dissipative and cools via dark bremsstrahlung (and
potentially 
other processes), which means that the cooling rate at the position 
${\bf r}$ is proportional to the square of dark matter density:
$\Gamma_{cool}({\bf r})  \propto n({\bf r})^2$. 
[The proportionality 
coefficients depend on dissipative particle physics and 
do not need to be specified for the present discussion.]\footnote{
The emitted dark photon radiation 
from a local volume
can only be an effective cooling agent if the halo is optically thin.
In specific models including mirror dark matter, the halo may in fact be 
optically thick for a range of dark photon wavelengths.
In which case we can consider as an approximation only the optically
thin
cooling component, as the optically thick component will be reabsorbed
and
to first order can be neglected.} 
Assuming now for simplicity an approximately isothermal halo\footnote{
The halo could not be exactly isothermal as this would be incompatible
with 
the hydrostatic equilibrium condition. Thus, corrections to the results 
following from the simple assumptions adopted here are in fact
inevitable.}, 
the dark matter density can be obtained directly from the presumed
dynamically 
driven local balancing of the heating and cooling rates,  
$\Gamma_{heat}({\bf r}) = \Gamma_{cool}({\bf r})$, so that: 
\begin{eqnarray}
n({\bf r}) \propto F_{\gamma_D} ({\bf r}).
\label{sunx}
\end{eqnarray}
As already mentioned, the timescale of this halo evolution is assumed to
be 
much smaller than the current age of the Universe. Given that the dark
photon 
energy flux is presumed to originate from ordinary supernovae, it can be 
straightforwardly calculated by summing up all the sources in the disk.
%%% new stuff herexxx%%
Define a spherical co-ordinate system with the origin at the center of
the 
galaxy and with the baryonic disk described in terms of 
$\widetilde{r}, \ \widetilde{\phi}$ at $\theta = \pi/2$.
If supernovae occur at an average rate per unit area of 
$\Sigma_{SN}(\widetilde{r},\widetilde{\phi})$, then
the flux of dark photons in a particular wavelength range from a disk
area 
element: $\widetilde{r}d\widetilde{\phi} d\widetilde{r}$ 
is proportional to: 
\begin{eqnarray}
dF_{\gamma_D}
\propto \frac{e^{-\tau} \ \Sigma_{SN}(\widetilde{r},\widetilde{\phi})}
{4\pi d^2} \ \widetilde{r}d\widetilde{\phi}  d\widetilde{r}\ . 
\end{eqnarray}
where $\tau$ is the wavelength dependent optical depth and $d$ is the
distance 
between the point ${\bf r}$ and the disk element 
defined by $\widetilde{r}, \widetilde{\phi}$.
%%%%
For an optically thin halo, this leads to the relatively simple formula
for 
the dark matter density profile:\footnote{As briefly noted already, the
halo 
may be optically thick for a range of wavelengths:  
$\lambda_L  \lesssim{<} \lambda_{\gamma_D}  \lesssim{<} \lambda_H$ 
(the precise values of $\lambda_L, \lambda_H$ are
model/parameter dependent and are uncertain). 
Nevertheless, the optically thin dark photons can dominate 
the energy transport if the energy spectrum of dark photons originating
in the 
region around ordinary supernova peaks at a wavelength below the 
low-wavelength cutoff scale, $\lambda_L$ \cite{10,11}.}
\cite{13}
\begin{eqnarray}
\rho(r,\theta,\phi) =  
\lambda \int
d\widetilde{\phi} \int d\widetilde{r} \ \widetilde{r} 
\ \frac{\Sigma_{SN}
({\widetilde{r},\widetilde{\phi}})} {4\pi[r^2 + {\widetilde{r}}^2 - 
2r\widetilde{r}  \sin\theta \cos
(\widetilde{\phi}-\phi)]}
\ .
\label{3z}
\end{eqnarray}
%Here, we have assumed a spherical co-ordinate system 
%with the baryonic disk located at $\theta = \pi$./

The proportionality coefficient $\lambda$ in Eq.(\ref{3z}) is a
combination 
of many quantities. In specific dissipative dark matter models it
depends on
fundamental parameters related to the dark photoionization cross section
and 
radiative cooling cross sections. It also depends on other quantities
including
the supernovae dark photon energy spectrum, and halo properties:
ionization 
state and composition. In general, we also expect $\lambda$ to depend 
on halo temperature and, hence, also on galaxy properties (mass,
luminosity 
etc) and therefore it is not expected to be a universal constant but can
have 
some (possibly weak) scaling between galaxies. Furthermore, $\lambda$
may not 
even be a constant within a galaxy but have dependence on spatial
coordinates. 
We assume that such variation is subleading. Some work 
justifying this assumption has been done \cite{10,11}, and more work is 
in progress \cite{wp} on this important issue for these kinds of dark matter
models. If these assumptions are indeed valid, then dissipative dark
matter 
can be extremely predictive as far as galaxy dynamics is concerned. 

\section*{Radial acceleration relation in dissipative dark matter model}
 
In order to explore the radial acceleration relation in the context
of dissipative dark matter,   
consider a generic disk galaxy with the supernova 
distribution approximated as an axisymmetric thin Freeman disk
\cite{Fdisk}:
\begin{eqnarray}
\Sigma_{SN}(r) = R_{SN} \frac{e^{-r/r_D}}{2\pi r_D^2}
\ ,
\label{disk}
\end{eqnarray}
where $R_{SN}$ is the type-II supernova rate in the galaxy under
consideration  and
$r_D$ is the disk scale length.

For an actual galaxy this exponential distribution could only be a rough 
approximation; a better approximation would be to use the measured UV
surface 
brightness profile of the galaxy under consideration \cite{14}. Anyway,
with 
the axisymmetric distribution given in Eq.(\ref{disk}), the dark matter 
density in Eq.(\ref{3z}) depends only on $r$ and $\theta$, and the dark
matter
contribution to the gravitational acceleration at a point in the plane
of 
the disk can be straightforwardly obtained from Newton's law of gravity: 
\begin{eqnarray}
g_{dark} (r) = G_N \int d\widetilde{\phi} \int d\cos\widetilde{\theta} 
\int d\widetilde{r} \ \widetilde{r}^2 \ {\rho (\widetilde{r},
\widetilde{\theta}) \cos\omega \over 
d^2}
\ .
\label{4z}
\end{eqnarray}
Here, $d^2 \equiv r^2 + {\widetilde{r}}^2 - 2r \widetilde{r}  
\sin\widetilde{\theta} \cos \widetilde{\phi}$,   
$\cos\omega \equiv (r - \widetilde{r}\sin\widetilde{\theta}
\cos\widetilde{\phi})/d$ and $G_N$ is Newton's constant. The baryonic 
contribution to the gravitational acceleration will also be needed, and
for 
an axisymmetric thin disk of stellar mass $m$ and surface density 
$\Sigma (\widetilde{r}) = m \ e^{-\widetilde{r}/r_D}/(2\pi r_D^2)$ it is 
described by an equation of the form:
Eq.(\ref{4z}) with $\rho(\widetilde{r},\widetilde{\theta}) \to \Sigma \ 
\delta(\widetilde{\theta} - \pi/2)/\widetilde{r}$.\footnote{For the
Freeman 
disk the gravitational acceleration in the equatorial plane
($\theta=\pi/2$)
is known analytically \cite{Fdisk}:
$g_B(x)= [G_N m_{bar}/r_D^2] 
\ (x/2)[ I_0(x/2) 
K_0(x/2) -I_1(x/2)K_1(x/2) ]$,
where $I_0,I_1$ are modified Bessel functions of the first kind,
$K_0,K_1$ 
are modified Bessel functions of the second kind, and $x \equiv r/r_D$.}

The dark matter density function motivated by dissipative dark matter, 
Eq.(\ref{3z}), is relatively simple and depends on only one parameter if 
the supernovae distribution is known.  
%For an exponential $\Sigma_{SN}$ profile with scale 
%length $r_D$ [Eq.(\ref{disk})], 
In the axisymmetric case with the exponential $\Sigma_{SN}$ profile
it is possible to do the $\widetilde{\phi}$-integration analytically. 
The result is that the density takes the form of a Laplace
transformation:
\begin{equation}
\rho(r,\theta)=
%\frac{\lambda R_{SN}}{4\pi r_D^2}\int\limits_0^\infty
%\frac{ye^{-y} dy}{\sqrt{r^4+y^4+2r^2y^2\cos{2\theta}}}=
%\frac{\lambda R_{SN}}{4\pi r_D^2}\int\limits_0^\infty
%\frac{\tilde re^{-\tilde r/r_D} d\tilde r}{\sqrt{r^4+\tilde r^4+
%2r^2\tilde r^2\cos{2\theta}}}=
\frac{\lambda R_{SN}}{4\pi r_D^2}\int\limits_0^\infty
\frac{te^{-rt/r_D} \ dt}{\sqrt{1+t^4+2t^2\cos{2\theta}}} \ .
\end{equation} 
%where at the final step we have made a substitution $\tilde r=rt$.
From this expression,
the radial dependence of the density 
for a fixed angular direction $\theta \neq \pi/2$
can be shown to satisfy:
\begin{eqnarray}
    \rho(r) = \left\{
                \begin{array}{ll}
\frac{\lambda R_{SN} |log (r/r_D)|}{4 \pi r_D^2} \ \ \ {\rm for} 
\ \ r \ll r_D  \\ 
\frac{\lambda R_{SN}}{4\pi r^2} \ \ \ {\rm for}  \ \ r \gg r_D \\
                \end{array}
              \right.
\label{mon}
\end{eqnarray}
The transition region occurs roughly when $r/r_D \approx 1.5$.
The density has a log divergence as $\theta \to \pi/2$ which would be 
regulated by considering a disk of finite thickness. The gravitational 
acceleration, however, is finite and thus a thin disk is suitable for 
the current purposes. Despite its angular dependence, such a density
profile 
results in rotation curves similar  to that from a spherically symmetric 
cored isothermal profile: $\rho = \rho_0 r_0^2/(r^2 + r_0^2)$ with 
$r_0 \approx 1.4 r_D$.

At low $r \lesssim r_D$ the approximately flat 
density profile [Eq.(\ref{mon})] implies a linearly rising rotation
curve: 
\begin{eqnarray}
v(r) \simeq  r \sqrt{2\pi G_N \lambda \Sigma_{SN}(0)/3}
%  \ \ \ {\rm for }  \ \ r \stackrel{<}{\sim} r_D
\ .
\end{eqnarray}
Such linearly rising rotation curves are seen in dwarf disk galaxies,
where 
dark matter generally dominates over ordinary matter even at low radii 
\cite{nat,rot4} (for recent studies see e.g. \cite{rot2}). In particular
note 
that the inner rotation curve slope satisfies: 
$dv/dr \propto \sqrt{\Sigma_{SN}(0)}$, that is, the rotation curve
slope 
is expected to scale with the square root of the central surface
brightness 
(in the UV band), which is in fact in agreement with observations
\cite{s2} 
(see also \cite{r2,lelli2} for related discussions). 
Considering next the $\rho \propto 1/r^2$ behaviour at large radii, 
$r \gg r_D$, Newton's laws imply that the rotation curve has a flat 
(i.e. radially independent) asymptotic rotational velocity, 
$v_{asym} = \sqrt{G_N \lambda R_{SN}}$,
in agreement with long standing 
observations \cite{rot0a,rot0b}.

The baryonic Tully Fisher relation ($m_{bar} \propto v_{asym}^4$)
\cite{btf} 
will require the approximate scaling: $\lambda R_{SN} \propto
\sqrt{m_{bar}}$.
Core-collapse supernovae have been observed in the local Universe to
occur with a rate that roughly matches this scaling \cite{snscaling}, 
which means that an approximately constant $\lambda$ (or modestly
varying $\lambda$) would suffice to reproduce the baryonic Tully Fisher 
relation.\footnote{Some work has been done attempting to determine the
scaling 
behaviour of $\lambda$ within the mirror dark matter context
\cite{10,11}. 
Although some simplifying assumptions were made, that work indicates
that 
$\lambda$ does indeed scale modestly, and $\lambda R_{SN}$ appears to
have 
a scaling consistent with observations. Further work is underway which
aims 
to provide a more rigorous check of this conclusion.}
Actually, this kind of dark matter model motivates a somewhat 
different form to the Tully Fisher-type scaling. Instead of converting
$\lambda R_{SN}$ into $m_{bar}$, which itself can only be obtained
indirectly 
as observations measure light not mass, one can use the luminosity in
the UV 
band [$L_{UV}$] as a proxy for the current star formation rate. Thus we
expect 
$R_{SN} \propto L_{UV}$, and hence $v_{asym} = \sqrt{G_N \lambda
R_{SN}}$
suggests that: $\lambda L_{UV} \propto v^2_{asym}$.
One can further motivate a $\lambda \propto 1/v_{asym}$ scaling in
simple 
dissipative dark matter models if bremsstrahlung effectively dominates
the 
cooling leading to the rough scaling: $L_{UV} \propto v_{asym}^3$ 
\cite{14}. 
%This Tully Fisher-like relation was compared in \cite{14} with THINGS
%spirals \cite{things} and LITTLE  
%dwarfs \cite{rot2} and found to be approximately valid.

Another interesting feature of the density profile Eq.(\ref{3z}) is that
it leads to gravitational accelerations that have scale invariance in 
a similar sense as MOND. That is, 
under the scale transformation $(t,\vec{r})\to \Lambda\,(t,\vec{r})$, 
$r_D \to \Lambda r_D$ (with $G_N, R_{SN}, \lambda$ unchanged) the
density
scales:  $\rho \to \rho/\Lambda^2$, and $g_{dark} \to g_{dark}/\Lambda$. 
A consequence of this is that the dark halo contribution to the
rotational 
velocity is scale invariant, this means that it will depend only on the 
dimensionless ratio $r/r_D$ rather than $r$ and $r_D$ separately. 
Such scale invariant dynamics is supported by observations, see e.g.
\cite{s3} 
for a recent discussion. This scale invariance feature, along with the 
baryonic Tully Fisher scaling ($\lambda R_{SN} \propto \sqrt{m_{bar}}$),
imply that the rotation curves have a characteristic form. The dark
matter 
rotational velocity function $v(r)$ for a particular value of $r_D, \
m_{bar}$ 
can be mapped onto any other $r_D, \ m_{bar}$ value.

\begin{figure}[t]
\centering
\includegraphics[width=0.48\linewidth,angle=270]{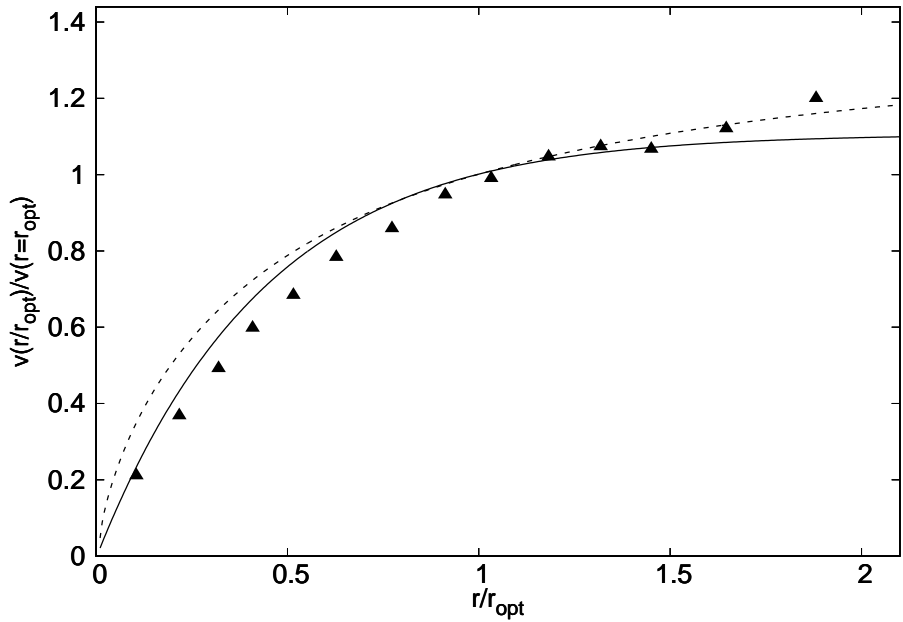}
\caption{The ratio $v(r/r_{opt})/v(r=r_{opt})$ versus $r/r_{opt}$. 
The solid line is the result for dissipative dark matter,  
i.e. from  $v^2/r = g_{dark}$ [Eqs.(\ref{4z}),(\ref{3z})]. 
The  curve is not dependent on any parameters (such as $r_D, 
\ m_{bar}$ or theory parameter $\lambda$). Also shown (dashed line) is 
the curve predicted by the radial acceleration relation in the deep-MOND 
regime where $v^2/r = \sqrt{a_0 g_{bar}}$. The triangles are the
synthetic 
rotation curve obtained from dwarf disk galaxies  \cite{sal2016}.}
\label{Fig1}
\end{figure}

The characteristic form for $v(r)$ can be conveniently specified by 
considering the velocity ratio (where, following \cite{sal2016}, the 
dimensionless parameter is taken to be $r/r_{opt}$ where the optical
radius
$r_{opt} \simeq 3.2 r_D$; although any other scaling with $r_D$ could be
used):
\begin{eqnarray}
R(r/r_{opt}) \equiv \frac{v(r/r_{opt})}{v(r=r_{opt})}
\ .
\label{r3}
\end{eqnarray}
Here the rotational velocity is that due to dark matter (i.e. obtained
from 
$v^2/r = g_{dark}$). It is worth emphasizing that 
with the exponential supernovae distribution profile chosen, the
velocity 
ratio Eq.(\ref{r3}) is predicted to be completely independent of the
particular value of galaxy parameters: $r_D, m_{bar}, R_{SN}$ as well as
the
theory parameter $\lambda$. This parameter-free theoretical curve 
shown in Fig.1, can be compared to actual rotation curves of dwarf disk 
galaxies as they are typically dark matter dominated at all radii
(summarized by synthetic rotation curves given in 
\cite{sal2016} and reproduced as the triangles in Fig.\ref{Fig1}).
Also shown for comparison is the curve predicted by the radial
acceleration 
relation in the deep-MOND regime where $v^2/r = \sqrt{a_0 g_{bar}}$.
This 
relation depends on the total baryonic gravitational field $g_{bar}$
which we 
have derived for a gas dominated dwarf with stellar/gas fraction of 
$f_{star} = 0.2$, $f_{gas} = 0.8$. (Both the stellar and gas components
are 
modelled with an exponential profile but the gaseous component 
is known from observations to be more radially extended which motivates
a 
larger scale radius for that component: $r_D^{gas} = 3.0*r_D$, e.g. 
\cite{sal2016}.)

%\vskip 0.5cm
%\centerline{\epsfig{file=fig1.eps,angle=270,width=11.7cm}}
%\vskip 0.2cm
%\noindent
%begin{caption}
%{\small
%Figure 1: The ratio $v(r/r_{opt})/v(r=r_{opt})$ versus $r/r_{opt}$. 
%The solid line is the dissipative dark matter value for the dark matter 
%contribution to this ratio, i.e. from  $v^2/r = g_{dark}$. 
%The  curve does not dependent on any free parameters (such as $r_D, 
%\ m_{bar}$ or theory parameters $\lambda$). Also shown (dashed line) is 
%the curve predicted by the radial acceleration relation in the
%deep-MONDian 
%regime where $v^2/r = \sqrt{a_0 g_{bar}}$. The triangles are the
%synthetic 
%rotation curve obtained from dwarf disk galaxies  \cite{sal2016}.
%}
%\vskip 0.9cm

\begin{figure}[ht]
  \begin{minipage}[b]{0.5\linewidth}
    \centering
    \includegraphics[width=0.7\linewidth,angle=270]{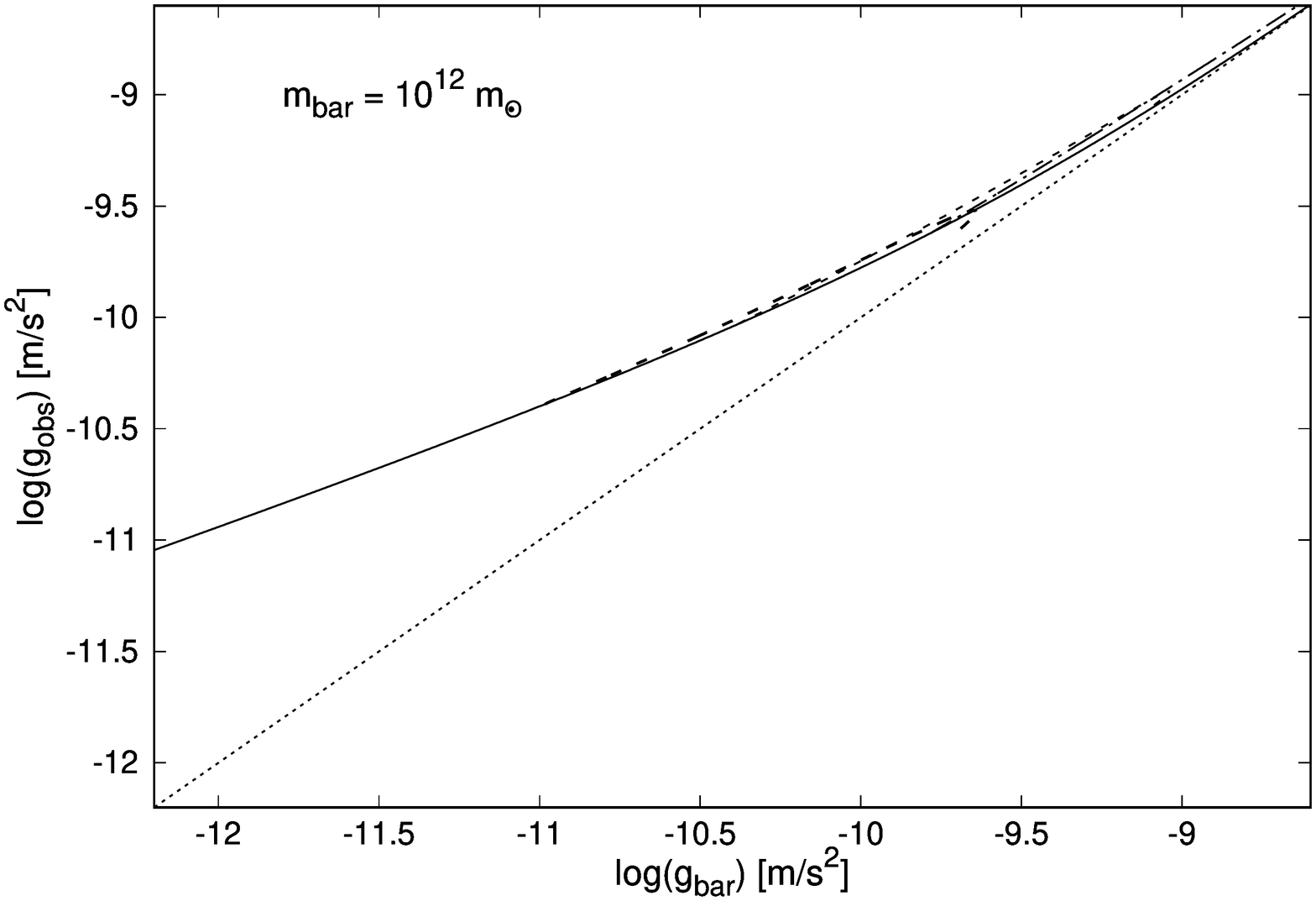}
     a)
    \vspace{4ex}
  \end{minipage}%%
  \begin{minipage}[b]{0.5\linewidth}
    \centering
    \includegraphics[width=0.7\linewidth,angle=270]{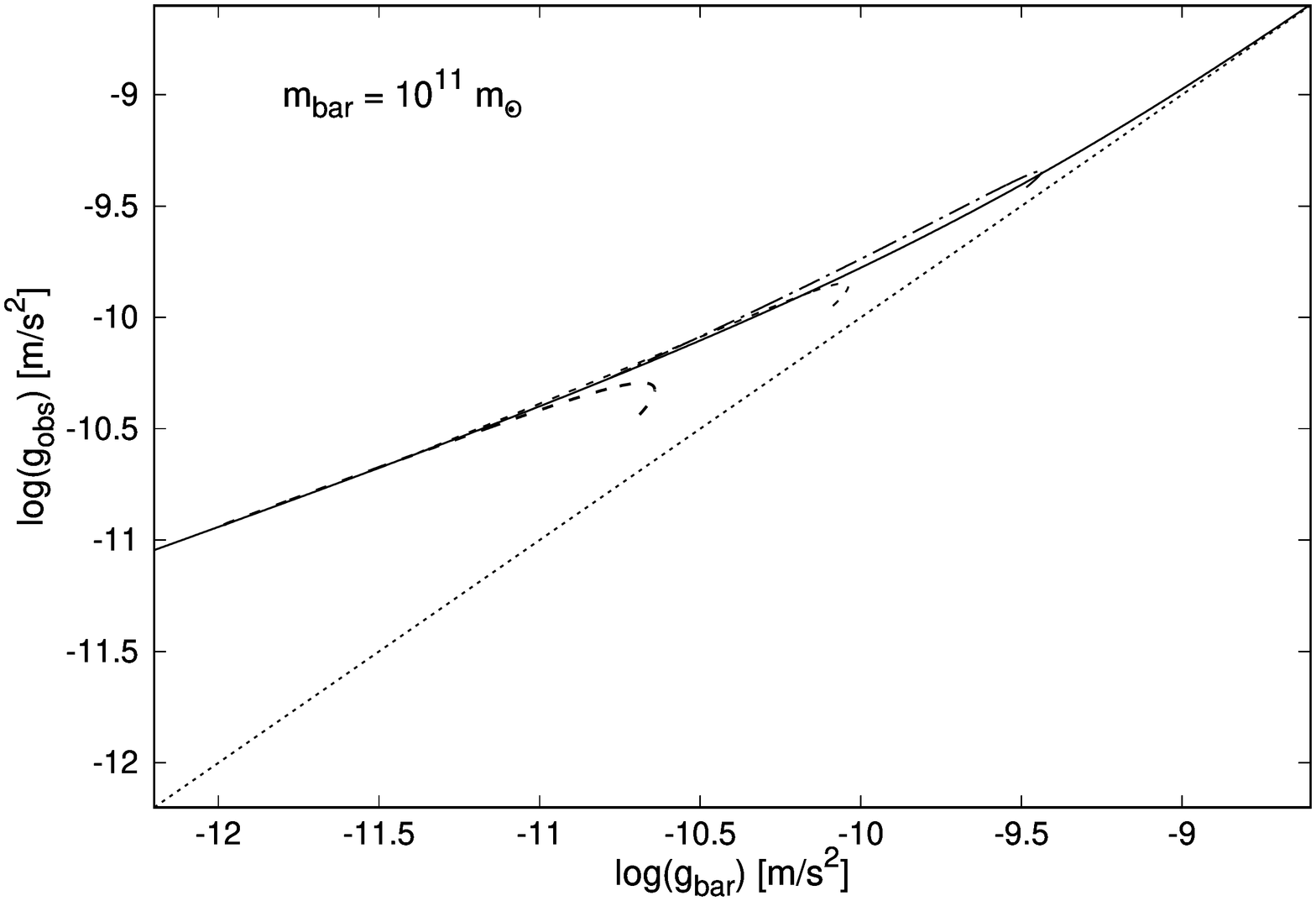}
    b)
    \vspace{4ex}
  \end{minipage}
  \begin{minipage}[b]{0.5\linewidth}
    \centering
    \includegraphics[width=0.7\linewidth,angle=270]{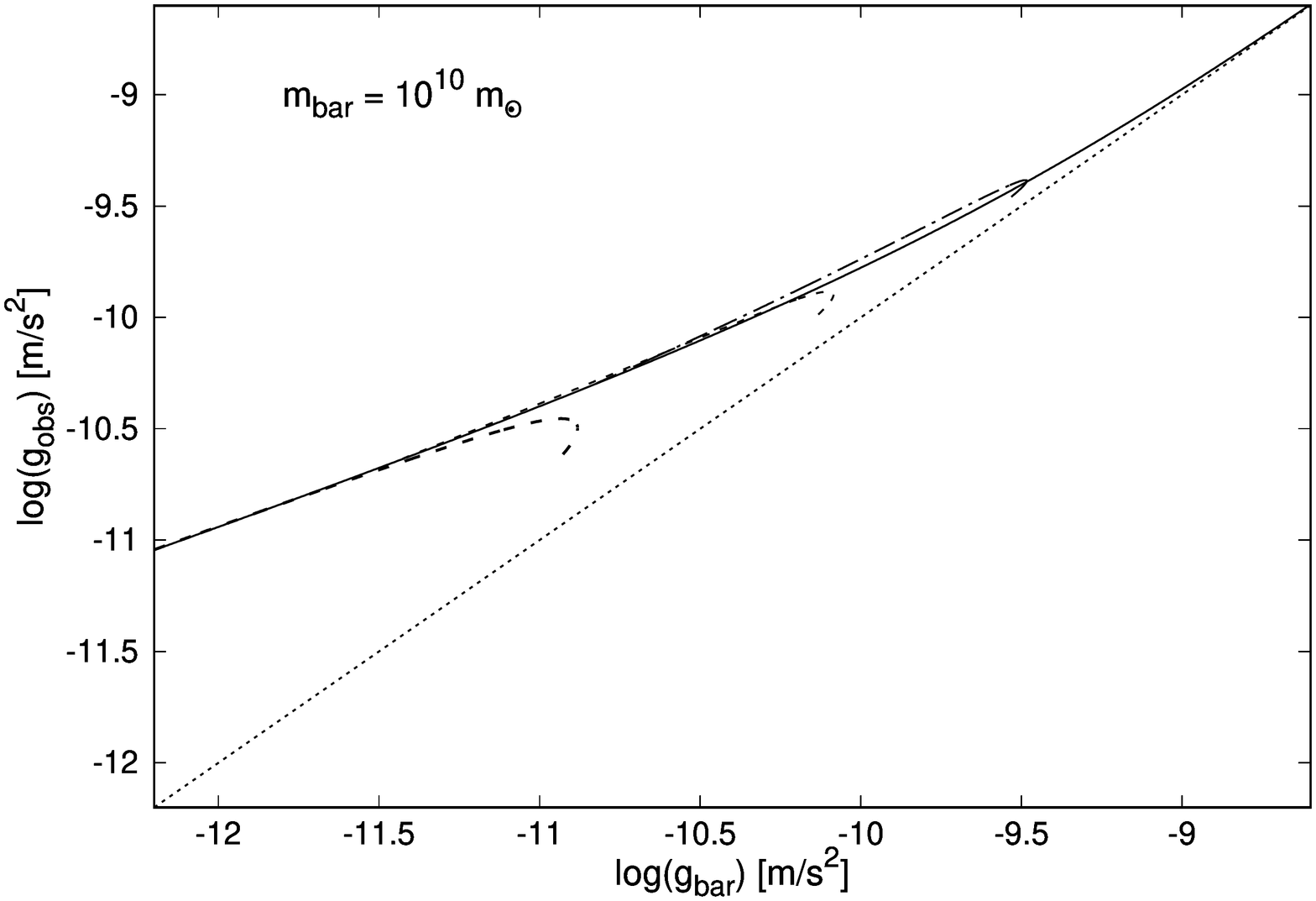}
    c)
    \vspace{4ex}
  \end{minipage}%%
  \begin{minipage}[b]{0.5\linewidth}
    \centering
    \includegraphics[width=0.7\linewidth,angle=270]{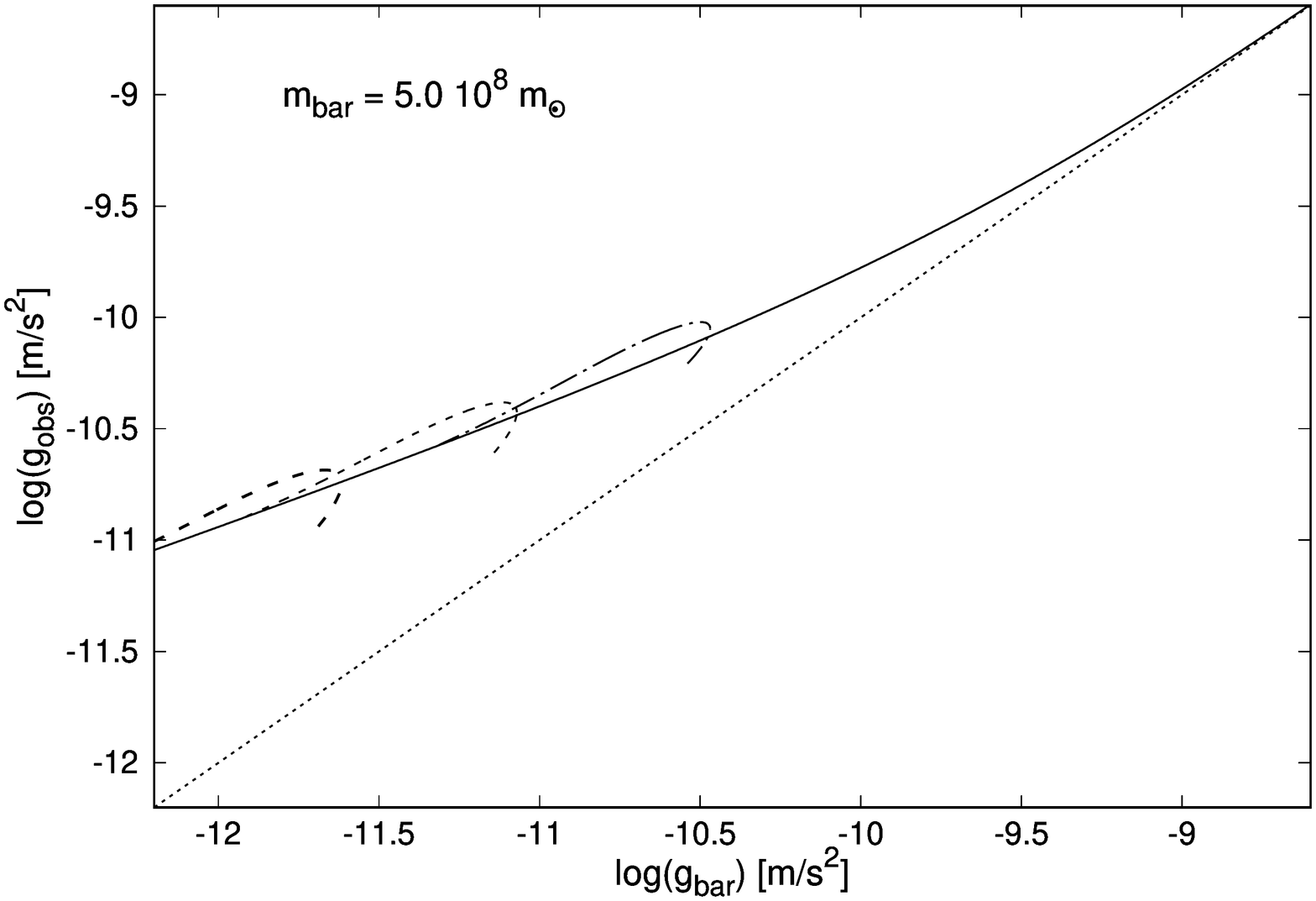}
    d) 
    \vspace{4ex}
  \end{minipage}
\caption{a) Radial acceleration relation, Eqs.(\ref{eq1}),(\ref{eq4})
[solid 
line], compared with dissipative dark matter, $g_{obs} =
g_{dark}+g_{bar}$, 
%[Eqs.(\ref{4z}),(\ref{3z})],
for illustrative examples. 
The exponential disk parameters chosen 
are $m_{bar}  = 10^{12} m_\odot$ and  $r_D = 3.0$ kpc (dashed-dotted
line), 
$r_D = 6.0$ kpc (thin dashed line) and $r_D = 12.0$ kpc (thick dashed
line). 
In each case a  stellar/gas mass fraction of $f_{star} = 0.8$, $f_{gas}
= 
0.2$ was used. Shown are accelerations obtained for $0.4 < r/r_D < 10$.
The dotted line is the $g_{obs} = g_{bar}$ limit. 
b) Same as Fig.2a, but with 
$m_{bar} = 10^{11} m_\odot$ and $r_D =3.0$ kpc (dashed-dotted line), 
$r_D = 6.0$ kpc (thin dashed line) and $r_D = 12.0$ kpc (thick dashed
line).
c) Same as Fig. 2a, but with 
$m_{bar}  = 10^{10} m_\odot$ and $r_D = 1.0$ kpc (dashed-dotted line),
$r_D = 2.0$ kpc (thin dashed line) and $r_D = 5.0$ kpc (thick dashed
line).
d) Same as Fig. 2a, but with $m_{bar}  = 5.0 \times 10^{8} m_\odot$ and 
$r_D = 0.4$ kpc (dashed-dotted line), $r_D = 0.8$ kpc (thin dashed line)
and 
$r_D = 1.5$ kpc (thick dashed line), with stellar/gas mass fraction of 
$f_{star} = 0.2$, $f_{gas}  = 0.8$.}
\label{Fig2}
\end{figure}

In Fig.\ref{Fig2} we give the radial acceleration relation predicted 
by the density profile Eq.(\ref{3z}) for some illustrative examples. 
The baryonic matter is modelled as a thin disk with both stellar and gas 
components each with an exponential profile (with  
$r_D^{gas} = 3.0*r_D$ as discussed above). 
Both the radial acceleration relation and the dark 
matter density motivated from dissipative dynamics have one free
parameter, 
and we have normalized $\lambda R_{SN}$ such that it gives the same
value for $g_{obs} = g_{dark} + g_{bar}$ as the $g$ obtained from the
formula, 
Eqs.(\ref{eq1}),(\ref{eq4}) (with $a_0= 1.20 \times
10^{-10}~\mathrm{m}\,
\mathrm{s}^{-2}$) for the largest radii considered.   
The radial acceleration relation is in agreement with observations to
within 
errors estimated to be 0.12 dex for the sample studied in \cite{8}.

Figures 2a, 2b, 2c  give our results for disk galaxies with $m_{bar}$
values 
in the range: $10^{10} m_\odot - 10^{12} m_\odot$. The examples assumed
a disk 
dominated by stars with stellar/gas mass fractions of $f_{star} = 0.8$, 
$f_{gas} = 0.2$ (typical for spiral galaxies). For each $m_{bar}$ value
we 
considered $r_D$ values indicative of high surface brightness spiral
galaxies, 
taken to be $r_D = 3.0$ kpc and $r_D  = 6.0$ kpc for $m_{bar} = 10^{11}
- 
10^{12} \ m_\odot$ and $r_D = 1.0$ kpc, $r_D = 2.0$ kpc for $m_{bar} =
10^{10} 
\ m_\odot$ (consistent with recent studies \cite{size}). We also
considered 
examples with larger $r_D$ values relevant to low surface brightness
galaxies.
In figure 2d we considered the radial acceleration relation for the
example of 
a smaller gas rich dwarf galaxy. The parameters chosen were: $m_{bar}  = 
5.0\times 10^8 \ m_\odot, r_D = 0.4$ kpc, $r_D = 0.6$ kpc, and $r_D =
1.5$ kpc,
with stellar/gas mass fractions  of $f_{star} = 0.2$, $f_{gas} = 0.8$ 
consistent with typical values for gas rich dwarfs.

The figures demonstrate that the density profile Eq.(\ref{3z}) motivated
by 
dissipative dark matter leads to accelerations within galaxies 
consistent with  
the radial acceleration relation, which itself is known 
to be a good representation of the data. The differences are small and
occur
mainly at  low radii $r \lesssim r_D$ where uncertainties are generally
larger 
(especially in spirals). The reasons for this agreement are several. 
First, as already discussed, the concept of scale invariance can be used
to explain the lack of significant $r_D$ variation between these curves.
If they approximately agree for one value of $r_D$ then they should 
approximately agree for any other. Second, the basic velocity profiles
in the 
dark matter dominated and deep-MONDian limits are very similar 
(Fig.\ref{Fig1}). We don't have any deep theoretical explanation for why
this 
results, other than the basic velocity profile is a function of the
baryonic 
distribution in each case. Finally, the agreement assumes $\lambda
R_{SN}$ 
values such that $g_{obs} = g_{dark} + g_{bar}$ is equal to the value of
$g$ 
obtained from the radial acceleration relation
[Eqs.(\ref{eq1}),(\ref{eq4})], 
at large radii. This is approximately equivalent to having the scaling: 
$\lambda R_{SN} \propto \sqrt{m_{bar}}$. As briefly mentioned already,
the 
galactic supernova rate itself has an observed scaling consistent with 
$R_{SN} \propto \sqrt{m_{bar}}$, and thus a roughly constant or weak
scaling  
of the model parameter $\lambda$ is required which appears to be
possible in 
the specific models studied \cite{10,11,wp}.

Previous work \cite{13,14} along with the present results indicate that 
Eq.(\ref{3z}) provides a reasonably successful quantitative description
of the physical properties of dark matter in disk galaxies.  We
emphasize 
again here that the dark matter density given by Eq.(\ref{3z}) could 
only be a rough approximation to the actual dark matter density as
dictated 
by the dissipative dynamics. As briefly mentioned earlier, a more exact 
description is given by Euler's equations of fluid dynamics. More work
is 
needed to solve these fluid equations in both the general time-dependent
case 
(applicable to e.g. star burst galaxies) as well as the simpler steady
state
solution applicable to isolated galaxies with stable star formation
rates.

\section*{Conclusion}

Observations have shown that the structural properties of
dark 
matter and baryons in galaxies are  deeply entwined. The radial
acceleration 
relation, which can be viewed as a summary of much of the relevant 
information, gives a simple analytic form to the apparent collusion.
It is an approximate empirical law that dark matter theories need to
explain 
if they are to describe nature. Dissipative dark matter models have the 
pertinent feature that they actually require baryons to influence dark
matter 
properties as ordinary core-collapse supernovae appear to be the only
viable 
heat source which can dynamically balance the radiative cooling of dark
matter 
halos. Dissipative dark matter thereby motivates a particular dark
matter 
density profile, Eq.(\ref{3z}), which, as we have shown here,
approximately reproduces the empirical radial acceleration relation.

\section*{Acknowledgments}
 The work of Z.K.S. is supported by the Ministry of Education 
and Science of the Russian Federation, while that of R.F. was supported
by 
the Australian Research Council.
We also thank F. Petrov for some helpful insight into the derivation of
the limit, Eq.(\ref{mon}).

\end{document}